\documentclass[twocolumn,reprint,showpacs, superscriptaddress, amsmath,amssymb, aps, prl]{revtex4-1}

\usepackage{graphicx}
\usepackage{amsmath}
\usepackage{epsfig}
\usepackage{color}
\usepackage[english]{babel}
\usepackage{dcolumn}
\usepackage{bm}
\usepackage[
   bookmarks=true,
   colorlinks=true,
   hyperindex=true,
   citecolor=blue,
   linkcolor=blue,
   urlcolor=blue
]{hyperref}

\begin{document}

\title{High Fidelity Single-qubit Gates of a Single Neutral Atom in the Magic-Intensity Optical Dipole Trap}

\author{Cheng Sheng}
\affiliation{State Key Laboratory of Magnetic Resonance and Atomic and Molecular Physics, Wuhan
Institute of Physics and Mathematics, Chinese Academy of Sciences - Wuhan National
Laboratory for Optoelectronics, Wuhan 430071, China}
\affiliation{School of Physics, University of Chinese Academy of Sciences, Beijing 100049, China}
\affiliation{Center for Cold Atom Physics, Chinese Academy of Sciences, Wuhan 430071, China}
\author{Xiaodong He}
\email{hexd@wipm.ac.cn}
\affiliation{State Key Laboratory of Magnetic Resonance and Atomic and Molecular Physics, Wuhan
Institute of Physics and Mathematics, Chinese Academy of Sciences - Wuhan National
Laboratory for Optoelectronics, Wuhan 430071, China}
\affiliation{Center for Cold Atom Physics, Chinese Academy of Sciences, Wuhan 430071, China}
\author{Ruijun Guo}
\affiliation{State Key Laboratory of Magnetic Resonance and Atomic and Molecular Physics, Wuhan
Institute of Physics and Mathematics, Chinese Academy of Sciences - Wuhan National
Laboratory for Optoelectronics, Wuhan 430071, China}
\affiliation{School of Physics, University of Chinese Academy of Sciences, Beijing 100049, China}
\affiliation{Center for Cold Atom Physics, Chinese Academy of Sciences, Wuhan 430071, China}

\author{Kunpeng Wang}
\affiliation{State Key Laboratory of Magnetic Resonance and Atomic and Molecular Physics, Wuhan
Institute of Physics and Mathematics, Chinese Academy of Sciences - Wuhan National
Laboratory for Optoelectronics, Wuhan 430071, China}
\affiliation{School of Physics, University of Chinese Academy of Sciences, Beijing 100049, China}
\affiliation{Center for Cold Atom Physics, Chinese Academy of Sciences, Wuhan 430071, China}
\author{Peng Xu}
\affiliation{State Key Laboratory of Magnetic Resonance and Atomic and Molecular Physics, Wuhan
Institute of Physics and Mathematics, Chinese Academy of Sciences - Wuhan National
Laboratory for Optoelectronics, Wuhan 430071, China}
\affiliation{Center for Cold Atom Physics, Chinese Academy of Sciences, Wuhan 430071, China}
\author{Zongyuan Xiong}
\affiliation{State Key Laboratory of Magnetic Resonance and Atomic and Molecular Physics, Wuhan
Institute of Physics and Mathematics, Chinese Academy of Sciences - Wuhan National
Laboratory for Optoelectronics, Wuhan 430071, China}
\affiliation{Center for Cold Atom Physics, Chinese Academy of Sciences, Wuhan 430071, China}
\author{Min Liu}
\affiliation{State Key Laboratory of Magnetic Resonance and Atomic and Molecular Physics, Wuhan
Institute of Physics and Mathematics, Chinese Academy of Sciences - Wuhan National
Laboratory for Optoelectronics, Wuhan 430071, China}
\affiliation{Center for Cold Atom Physics, Chinese Academy of Sciences, Wuhan 430071, China}
\author{Jin Wang}
\affiliation{State Key Laboratory of Magnetic Resonance and Atomic and Molecular Physics, Wuhan
Institute of Physics and Mathematics, Chinese Academy of Sciences - Wuhan National
Laboratory for Optoelectronics, Wuhan 430071, China}
\affiliation{Center for Cold Atom Physics, Chinese Academy of Sciences, Wuhan 430071, China}
\author{Mingsheng Zhan}
\email{mszhan@wipm.ac.cn}
\affiliation{State Key Laboratory of Magnetic Resonance and Atomic and Molecular Physics, Wuhan
Institute of Physics and Mathematics, Chinese Academy of Sciences - Wuhan National
Laboratory for Optoelectronics, Wuhan 430071, China}
\affiliation{Center for Cold Atom Physics, Chinese Academy of Sciences, Wuhan 430071, China}
\date{\today}

\begin{abstract}
We demonstrate high fidelity single-qubit gate operation in a trapped single neutral atom. 
The atom is trapped in the recently invented magic-intensity optical dipole trap (MI-ODT) with more stable magnetic field. 
The MI-ODT efficiently mitigates the detrimental effects of light shifts thus sufficiently improves the performance of single qubit-gates.
The gates are driven with microwave, and the fidelity of gate operation is characterized by using the randomized benchmarking method. 
We obtain an average error per Clifford gate of $3.0(7)\times10^{-5}$ which is much below the error threshold ($10^{-4}$) for fault-tolerance. 
This error is found to be dominated by qubit dephasing, and the corresponding coherence time relevant to the Clifford gates is also measured experimentally.
This work is an essential step toward the construction of a scalable quantum computer with neutral atoms trapped in an MI-ODT array.
\end{abstract}



\maketitle


Due to the good scalability of neutral atoms trapped in a controllable large optical dipole trap (ODT) arrays, neutral atom array becomes one of the important quantum systems for building a quantum computer and quantum simulator~\cite{saffman2010, Georgescu2014,Weiss2017,Bloch2017}.
Recent experimental progress has demonstrated atom-by-atom assemblers of defect-free arbitrary 1D and 2D atomic arrays with up to 50 single atoms~\cite{Endres2016,Barredo2016}. Furthermore, two recent experiments have provided detailed characterization of microwave-driven single-qubit gate fidelities at Stark shift selected sites in 49-qubit 2D array and 125-qubit 3D arrays, respectively~\cite{Xia2015,Wang2016}. These progresses are important steps along the path of converting the scalability promise of neutral atoms into reality.

Principally, the performance of quantum computers relies on quantum gates with errors small enough to enable fault-tolerant operation through the use of quantum error correction protocols. Normally, the required maximum tolerable error varies between correction strategies, the commonly accepted error threshold per gate is 10$^{-4}$~\cite{Knill2005,Steane1996,Devitt2013}. To date, single-qubit gates with errors below this threshold have been achieved in many other well-known physical qubits, including trapped ions\cite{Brown2011,Harty2014}, superconducting qubits~\cite{Barends2014},quantum dots\cite{Veldhorst2014} and NV center\cite{Rong2014}. But in neutral-atom qubits, the achievement of single-qubit gates with error below the above error threshold remains elusive.

With a typical physical qubit encoded in hyperfine ground states of heavy alkali atoms (Rb,Cs), the single-qubit gates undergo large error that mainly arises from the so-called differential light shift (DLS) in previous experiments~\cite{Olmschenk2010,Xia2015,Wang2016}.
Generally, because of several GHz's hyperfine structure splitting, different hyperfine states of atomic qubit experience mismatched light shifts induced by the trapping laser field, leading to the detrimental DLS. This DLS depends on the laser intensity at the qubit position. Due to the spatial distribution of laser intensity in a trap, the qubit suffers from not only a strong inhomogeneous dephasing effect due to energy distribution of single atoms, but also the homogeneous dephasing caused by the intensity fluctuations and pointing instability of the ODT~\cite{Kuhr2005,Yu2013}. The former leads to uncontrollable frequency detuning up to over one-hundred Hz and causes frequency errors~\cite{Xia2015,Wang2016}, the latter limits the $1/e$ decay time denoted by $T_{2s}$ of the spin-echo visibility~\cite{Olmschenk2010}. Combined with relative slow microwave-driven Rabi frequency (10 kHz typically), all the perviously reported values of error per single-qubit gate are above the error threshold~\cite{Olmschenk2010,Xia2015,Wang2016}.

One of the solutions to the above problem is to construct a ``magic'' trap to completely control the DLS. Recently, we have constructed a magic-intensity ODT (MI-ODT) for $^{87}$Rb qubits with a circularly polarized trapping field~\cite{Yang2016}. In this new type of ODT, the detrimental effects of DLS are efficiently mitigated and the coherence time measured by Ramsey experiment (denoted by $T_{2R}$) has been greatly extended to over 200 ms. This magic trapping technique relies on the hyperpolarizability contribution to the light shifts, and has been proved to be efficient to mitigate coherence loss in manipulating the mobile qubits. In this letter, for the first time, we show that this newly developed magic trapping technique can also efficiently improve the performance of microwave-driven single-qubit gates with errors substantially below the error threshold for fault-tolerant quantum computer.

We begin by improving the MI-ODT to extend atomic qubit's coherence time through actively stabilizing the magnetic-field. Single-qubit Clifford gates are driven by microwave pulses with precisely controlled frequency, duration and phase. After precise calibration of resonant frequency and Rabi frequency of qubits, we perform randomized benchmarking (RB) to characterize the fidelity of Clifford gates on neutral-atom qubits confined in MI-ODT. The gate operation error sources are then searched and attributed.

A physical qubit is encoded into microwave clock state of $^{87}$Rb atom as $|0\rangle\equiv|F=1,m_F=0\rangle$ and $|1\rangle\equiv|F=2,m_F=0\rangle$.
The MI-ODT and main experimental details on qubit manipulation in MI-ODT are described in Ref.~\cite{Yang2016}. In brief, single atoms are loaded from magneto-optical trap into a tightly focused linearly polarized ODT with 900 nm beam waist. The working wavelength of the ODT is about 830 nm. Then $\pi$ polarized 795 nm light resonant with $|5S_{1/2},F=1\rangle\rightarrow|5P_{1/2},F'=1\rangle$ (D1 line) and $\pi$ polarized 780 nm repump light resonant with $|5S_{1/2},F=2\rangle\rightarrow|5P_{3/2},F'=2\rangle$ (D2 line) are applied to optically pump atoms to $|0\rangle$ in 3.2 G magnetic bias field. The linearly polarized ODT is then changed to a circularly polarized ODT (the MI-ODT) by a controlled liquid crystal retarder within 30 ms. The single-qubit in the MI-ODT is rotated by a microwave radiation at frequency near 6.834 GHz. Compared with our pervious work~\cite{Yang2016}, in present setup the microwave radiation has been improved a lot for precisely rotating the qubit. The required resonant microwave radiations are generated by a commercial 6.834 GHz rubidium frequency synthesizer (Spectra Dynamics, Model RB-1). Its resolution limits of microwave phase and frequency instability are 0.09 degree and 1 $\mu$Hz respectively. The synthesizer output passes through a transistor-transistor (TTL)-controlled solid state pin switch (SP123DHS-80), that creates approximately rectangular-shaped pulse. This pulse is then amplified and delivered to a microwave horn external to the vacuum glass cell. The measured standing-wave ratio of the horn is 1.15. To passively stabilize the microwave amplitude, both working temperatures of the pin switch and amplifier are stabilized within 0.05 mK. With 1 Watt of output microwave, the Rabi oscillation frequency is optimized up to about 12 kHz.

The magic-trapping for $^{87}$Rb atoms requires a magnetic field of several Gauss, so the qubit becomes sensitive to the noise of magnetic field. With the working B field of 3.2 G, the homogenous dephasing from magnetic field noise dominates the decay time of the Ramsey signal in our previous experiments~\cite{Yang2016}. This dephasing is irreversible and cannot be efficiently refocused by spin-echo technique, i.e., by applying an additional pulse between the two Ramsey $\pi$/2 pulses~\cite{Kuhr2005}. The coherence time $T_{2s}$ has been found to dominate the measurement error in neutral atomic qubits confined in an optical lattice~\cite{Olmschenk2010}. To reduce the magnetic field noise and to obtain longer $T_{2s}$, we implement an active magnetic-field stabilization~\cite{Dedman2007}. The short-term stability decreases from 4 mG to sub-mG, measured by using a fluxgate magnetometer. The recorded envelope of spin-echo visibility is plotted in Fig.1. The achieved longest coherence time is $T_{2s}\approx1.72(8)s$.

\begin{figure}[htbp]
\centering
\includegraphics[width=8cm]{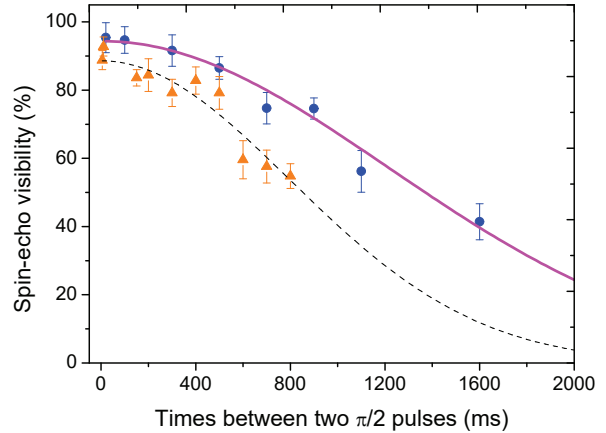}
\caption{(color online). The record longest coherence time $T_{2s}$. The data points ($\bullet$) are measured visibility of spin-echo as a function of time between two $\pi$/2 pulses with active magnetic-field stabilization. And in order to make a comparison, we plot the data points ($\blacktriangle$) of visibility measured without stabilization. All the accompanying error bars are fit uncertainties. The solid line and dashed line are fits to the Gaussian function yielding respectively coherence time $T_{2s}$ of 1.72(8)s and 1.13(7) s. }
\label{fig:fig1}
\end{figure}


To characterize the errors of single-qubit gates driven by the microwave radiation,
we adopt a well-developed RB method~\cite{Knill2008}. It has been proved to be a powerful technique that can distinguish gate errors from state preparation and measurement errors. We carry out the RB experiment by following the procedure described in Ref.~\cite{Xia2015}. In RB experiments, we use a complete set of 24 Clifford single-qubit gates $C_1$, which are generated from a set \{I,$R_j(\pm \pi/2)$,$R_j(\pi)$\} where I is the identity gate and $R_j(\theta)=e^{-\imath\theta\sigma_j/2}$ with $\sigma_j$ Pauli matrices about axes $j=x,y,z$. The basic spin rotation $R_j(\theta)=e^{-\imath\theta\sigma_j/2}$ are
from precisely control of frequency, duration and phase of microwave pulses.
Given a Rabi frequency $\Omega_M$, a resonant microwave pulse with time duration $t_d$ will rotate the single qubits in the Bloch sphere along the x-axis over an angle $\theta=\Omega_M t_d$ , that is $R_x(\theta)$ rotations. Shifting phase of the microwave pulse by a $\pi/2$ provides $R_y(\theta)$ rotations. The rotations about z-axis $R_z$ are implemented by composing $R_x$ and $R_y$ rotations. Since our frequency synthesizer can only provide phase shift keying modulation function, we replace all $R_{x,y}(-\pi/2)$ operations by $R_{x,y}(3\pi/2)$ rotations.

After preparing a qubit in $|0\rangle$ in the MI-ODT, random Clifford gate sequences of length $\ell$ for each gate are selected uniformly from $C_1$ and are implemented. At the end of each sequence, a final calculated Clifford gate is added and applied to bring the qubit state back to $|0\rangle$ with probability of 100$\%$ if errors would be absent. As the length $\ell$ increasing, the accumulated inevitable gate error will reduce transfer fidelity and the average probability of $|0\rangle$ state decays exponentially as follow~\cite{Knill2008}
\begin{equation}
\overline{F}=\frac{1}{2}+\frac{1}{2}(1-d_{if})(1-2\varepsilon_g)^\ell,
\end{equation}
where $d_{if}$ is the depolarization probability associated with
state preparation and $|0\rangle$ read-out, while $\varepsilon_g$ is the average error per Clifford gate. Thus, we can apply sequences of varying $\ell$ of randomized Clifford gates to single qubits initialized in $|0\rangle$ and fit the measured decay in average $|0\rangle$ fidelity to Eq. (1) to determine values of $d_{if}$ and $\varepsilon_g$.
\begin{figure}[htbp]
\centering
\includegraphics[width=8cm]{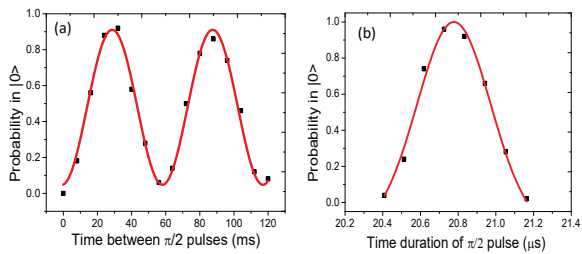}
\caption{(color online). Calibration of qubit's resonant frequency and Rabi frequency. (a)Measured Ramsey signals for single static qubits at B=3.2 G. Every point is an average over 50 experimental runs. The solid curves are fits to the sinusoidal function. The fitted
values of oscillation frequency is about 17.0(2) Hz. (b)Measured probability in $|0\rangle$ as a function of the pulse duration. Every point is an average over 50 experimental runs. A Gaussian fit to the data gives a central value at 20.78(1) $\mu s$.}
\label{fig:fig2}
\end{figure}


To precisely control qubit transition frequency and pulse durations, we perform two corresponding calibrations. First, we implement high precision microwave Ramsey spectroscopy to calibrate the resonant frequency of the qubit. The Ramsey experiment is performed by deliberately detuning by about 15 Hz and record the fringes by varying the time between the two $\pi/2$-pulses. A sinusoidal fit to the resulting oscillation, shown as in Fig.2(a), with a delay time as long as 120 ms, yields the qubit transition frequency with fitted uncertainty less than 0.2 Hz.
Then we calibrate the pulse duration $t_d$ providing a $R_x(\pi/2)$ rotation. To obtain high sensitivity to pulse duration, a sequence of 100 identical microwave pulses are applied on the single qubits in state of $|0\rangle$, by scanning the time of single pulse to produce a highest probability of finding the atoms in $|0\rangle$ state, as shown in Fig.2 (b). The typical $\pi/2$ time is $t_d\approx$20.78 $\mu s$, with an uncertainty of about 0.01 $\mu s$ from a Gaussian fit.

After the experimental parameter calibrations, we apply 5 different randomized Clifford gate sequences, each sequence is truncated at 7 different lengths \{1, 200, 400, 600, 800, 1000, 1300\}. For every truncated length, each kind of sequence is implemented 50 times to obtain a probability in $|0\rangle$. The average fidelity of of the 5 randomized Clifford gate sequences at each truncation length is plotted in Fig.3. A fit of these data to Eq.(1) yields an average error per gate of $\varepsilon_g\approx3.0(7)\times10^{-5}$ and $d_{if}\approx$0.03$\pm0.01$. We note that during the whole RB experiments, we did not regularly recalibrate the qubit transition frequency and the pulse durations. This measurement result indicates that the magic trapping technique can  efficiently improve the performance of microwave-driven single-qubit gates with errors substantially below the error threshold ($10^{-4}$) for fault-tolerance.

\begin{figure}[htbp]
\centering
\includegraphics[width=6cm]{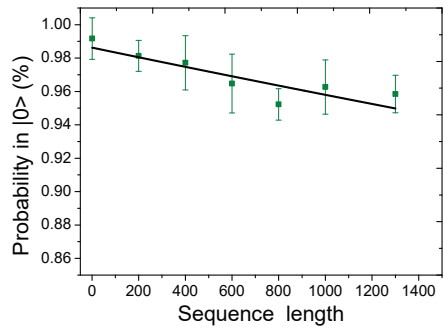}
\caption{(color online). Results of the single-qubit Clifford gate benchmarking experiment. Probability of measuring the $|0\rangle$ after applying 5 different random RB sequences truncated at different lengths $\ell$={1, 200, 400, 600, 800,1000, 1300}. The closed points are the average values of five different RB sequences at very value of $\ell$. Error bars are statistical, and
represent the standard deviation of the mean. The solid line is a fit to the Eq.(1) yielding respectively error per gate of $\varepsilon_g\approx3.0(7)\times10^{-5}$ and state preparation errors and readouts errors $d_{if}\approx0.03\pm0.01$.}
\label{fig:fig2}
\end{figure}


To understand the sources contributing to the above measured error per gate, we examine three main sources of error during RB experiment, (1) frequency detuning from the qubit transition; (2) fluctuations of $\pi/2$ pulse area; (3)
qubit dephasing. First we repeat the Ramsey experiment as shown in Fig.2(a) many times in a periods of 12 hours. The measured peak to peak is about 5 Hz, and the root-mean-square (rms) difference is about 1 Hz, which is mainly due to the drift of the magnetic field. This rms difference gives a predicted error per gate of 2.5$\times10^{-8}$, which obviously can be omitted.
In our experiment, pulse duration is controlled by a direct-digital synthesizer (DDS), which has a rms jitter of 200 ps. So the fluctuations of $\pi/2$ pulse area are mainly due to the pulse amplitude and microwave-atom coupling strength fluctuations. These fluctuations can be easily measured by repeating spectrum
shown as in Fig.2(b) for many times. After 3 hours' accumulation, we obtain the rms difference of pulse area of about 0.15$\%$. The corresponding error per gate is estimated to be $7.4\times10^{-6}$ per gate.

Apart from the above two minor effects, the contribution of dephasing to the error can be roughly estimated by taking the ratio of our qubit’s coherence time of $T_{2s}\approx$1.72 s to the average gate time of 75.73 $\mu s$, yielding an error per gate of $4.4\times10^{-5}$, which is larger than the measured value. This
disagreement suggests that the coherence time measured by spin-echo experiment is larger than the one at short time scale relevant to the single-qubit gate~\cite{Brown2011}, since the spin-echo experiments measure the decay of phase coherence for large magnitudes over long time scales.
We define a characteristic coherence time $T_{RB}$ to describe the phase noise at short time scale relevant to RB experiment, and assume that a ratio of $T_{RB}$ to $T_{2s}$ is $\eta$.

To determine $\eta$ in the RB experiment, we measure the dependence of errors $\varepsilon_g$ on coherence time $T_{2s}$, as shown in Fig.4. Here the coherence time $T_{2s}$ is scanned by increasing the laser intensity of ODT. Because of the parabolic intensity dependence of light shifts in a circularly polarized ODT, atoms experience more homogeneous dephasing effect when the working intensity is away from the magic point, as illustrated by the solid curves in inset of Fig.4. We can immediately see that under the same control pulse duration, noise fluctuations of magnetic field and microwave power, the error obviously increases as the coherence time reduces. Neglecting the contribution of pulse area error to the total measured errors, the $\varepsilon_g$ is given by
\begin{equation}
\varepsilon_g(T_{2s})=1 - \mathrm{e}^{-t_{CG}/{\eta*T_{2s}}},
\end{equation}
where $t_{CG}\approx$75.73 $\mu s$ is the average time duration of Clifford gates, including idle time of 3 $\mu s$. A fit to Eq.(2) yields $\eta\approx1.30(7)$.
From the fitting, we can see that, in our experiments, the short period phase noise can be scaled to the coherence time $T_{2s}$ and dominates the error per gate. This result is qualitatively in agreement with the one shown in the previous experiment done in optical lattice~\cite{Brown2011}.
As the results shown in Fig.4, we want to emphasise that on account of the $T_{RB}$ is robust to trap depth, we could implement a highly uniform global single-qubit gate with high fidelity in a MI-ODT array, that is hardly achieved otherwise in linearly polarized ODT array.

\begin{figure}[htbp]
\centering
\includegraphics[width=8cm]{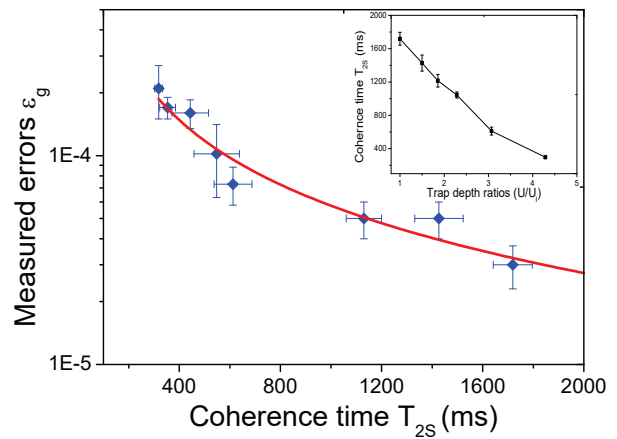}
\caption{(color online). Dependence of values of error per gate on the coherence times $T_{2s}$. All values of error per gate are measured by using RB method. All the accompanying error bars of coherence times and values of error per gate are fitting errors. The solid curve is fit to the Eq.(2). The inset shows dependence of coherence time $T_{2s}$ on ratio of varied trap depths (U) to magic working trap depth ($U_m$). The accompanying solid curve is to guide the eye.}
\end{figure}

In summary, we have experimentally proved that the novel MI-ODT can efficiently improve the performance of microwave-driven single-qubit gates with errors substantially below the error threshold for fault-tolerance. An average error per Clifford gate of $3.0(7)\times10^{-5}$ was realized in the updated MI-ODT. This error was found to be dominated by the qubit dephasing, and the characteristic coherence time  $T_{RB}$ was experimentally determined. This work, together with our previous demonstration of a coherent transfer of a mobile qubit~\cite{Yang2016} and recent demonstration of entangling two individual atoms of different isotopes ($^{87}$Rb and $^{85}$Rb) via Rydberg blockade~\cite{zeng2017}, provides another key element for a scalable quantum computer with neutral atoms trapped in an MI-ODT array.


\begin{acknowledgments}
This work was supported by the National Key Research and Development Program of China under Grant
Nos.2017YFA0304501, 2016YFA0302800, and 2016YFA0302002, the National Natural Science Foundation of China under Grant No.11774389, and the Strategic Priority Research Program of the Chinese Academy of Sciences under Grant No.XDB21010100.
\end{acknowledgments}


\begin{thebibliography}{99}
\bibitem{saffman2010} M. Saffman, T.G. Walker, and K. M\"{o}lmer, Quantum information with Rydberg atoms, Rev. Mod. Phys. {\bf 82}, 2313 (2010).
\bibitem{Georgescu2014}I.M. Georgescu, S. Ashhab, and F. Nori, Quantum simulation, Rev. Mod. Phys. {\bf86}, 153 (2014).
\bibitem{Weiss2017}D. S. Weiss and M. Saffman, Quantum computing with neutral atoms, Phys. Today {\bf 70}, 44 (2017).
\bibitem{Bloch2017} C. Gross and I. Bloch, Quantum simulations with ultracold atoms in optical lattices, Science, {\bf 357}, 995,(2017).
\bibitem{Endres2016}  M. Endres, H. Bernien, A. Keesling, H. Levine, E. R. Anschuetz, A. Krajenbrink, C. Senko, V. Vuletic, M. Greiner, and M. D. Lukin, Atom-by-atom assembly of defect-free one-dimensional cold atom arrays, Science {\bf 354}, 1024 (2016).
\bibitem{Barredo2016} D. Barredo, S. L\'{e}s\'{e}leuc, V. Lienhard, T. Lahaye, and A. Browaeys, An atom-by-atom assembler of defect-free arbitrary two-dimensional atomic arrays, Science {\bf 354}, 1021 (2016).
\bibitem{Xia2015} T. Xia, M. Lichtman, K. Maller, A. W. Carr, M. J. Piotrowicz, L. Isenhower, and M. Saffman, Randomized benchmarking of single-qubit gates in a 2D array of neutral-atom qubits, Phys. Rev. Lett. {\bf 114}, 100503 (2015).
\bibitem{Wang2016} Y. Wang, A. Kumar, T. Wu, and D. S. Weiss, Single-qubit gates based on targeted phase shifts in a 3D neutral atom array, Science {\bf 352}, 1562 (2016).
\bibitem{Knill2005} E. Knill, Quantum computing with realistically noisy devices, Nature (London) {\bf 434}, 39 (2005).
\bibitem{Steane1996} A. M. Steane, Error Correcting Codes in Quantum Theory, Phys. Rev. Lett. {\bf 77}, 793 (1996).
\bibitem{Devitt2013} S.J. Devitt, W.J. Munro, and K. Nemoto, Quantum error correction for beginners, Rep. Prog. Phys. {\bf 76}, 076001 (2013).
\bibitem{Brown2011}K.R. Brown, A.C. Wilson, Y. Colombe, C. Ospelkaus, A.M. Meier, E. Knill, D. Leibfried, and D.J. Wineland, Single-qubit-gate error below 10$^{-4}$ in a trapped ion, Phys. Rev. A {\bf 84}, 030303(R)(2011).
\bibitem{Harty2014}T.P. Harty, D.T.C. Allcock, C.J. Ballance, L. Guidoni, H.A. Janacek, N.M. Linke, D.N. Stacey, and D.M. Lucas, High-Fidelity Preparation, Gates, Memory, and Readout of a Trapped-Ion Quantum Bit, Phys. Rev. Lett. {\bf 113}, 220501 (2014).
\bibitem{Barends2014} R. Barends, J. Kelly, A. Megrant, A. Veitia, D. Sank, E. Jeffrey, T. C. White, J. Mutus, A. G. Fowler, B. Campbell, Y. Chen, Z. Chen, B. Chiaro, A. Dunsworth, C. Neill, P. O’Malley, P. Roushan, A. Vainsencher, J. Wenner, A. N. Korotkov, A. N. Cleland, and J. M. Martinis, Superconducting quantum circuits at the surface code threshold for fault tolerance, Nature {\bf508}, 500 (2014).
\bibitem{Veldhorst2014} M. Veldhorst, J.C. Hwang, C.H. Yang, A.W. Leenstra, B. de Ronde, J.P. Dehollain, J. T.Muhonen, F.E. Hudson, K.M. Itoh, A. Morello, and A.S. Dzurak, An addressable quantum dot qubit with fault-tolerant control-fidelity, Nat. Nanotechnol. {\bf9}, 981 (2014).
\bibitem{Rong2014}X. Rong, J.P. Geng, F.Z. Shi, Y. Liu, K.B. Xu, W.C. Ma, F. Kong, Z. Jiang, Y. Wu, and J.F. Du, Experimental fault-tolerant universal quantum gates with solid-state spins under ambient conditions, Nat. Commun. {\bf6},8748 (2015).

\bibitem{Olmschenk2010} S. Olmschenk, R. Chicireanu, K.D. Nelson, and J.V. Porto, Randomized benchmarking of atomic qubits in an optical lattice, New J. Phys. {\bf 12}, 113007 (2010).

\bibitem{Kuhr2005} S. Kuhr, W. Alt, D. Schrader, I. Dotsenko, Y. Miroshnychenko, A. Rauschenbeutel, and D. Meschede, Analysis of dephasing mechanisms in a standing-wave dipole trap, Phys. Rev. A.  {\bf 72}, 023406 (2005).
\bibitem{Yu2013} S. Yu, P. Xu, X. D. He, M. Liu, J. Wang, and M. S. Zhan, Suppressing phase decoherence of a single atom qubit with Carr-Purcell-Meiboom-Gill sequence, Opt. Express {\bf 21}, 32130(2013).
\bibitem{Yang2016}J. Yang, X.D. He, R.J. Guo, P. Xu, K.P. Wang, C. Sheng, M. Liu, J. Wang, A. Derevianko, and M.S. Zhan, Coherence Preservation of a Single Neutral Atom Qubit Transferred between Magic-Intensity Optical Traps, Phys. Rev. Lett. {\bf 117}, 123201 (2016).

\bibitem{Dedman2007}C.J. Dedman, R.G. Dall, L.J. Byron, and A.G. Truscott, Active cancellation of stray magnetic fields in a Bose-Einstein condensation experiment, Rev. Sci. Instrum. {\bf 78}, 024703 (2007).
\bibitem{Knill2008}E. Knill, D. Leibfried, R. Reichle, J. Britton, R. B. Blakestad,
J. D. Jost, C. Langer, R. Ozeri, S. Seidelin, and D. J. Wineland, Randomized benchmarking of quantum gates, Phys. Rev. A {\bf 77} 012307,(2008).

\bibitem{zeng2017}Y. Zeng, P. Xu, X.D. He, Y.Y. Liu, M. Liu, J. Wang, D.J. Papoular, G.V. Shlyapnikov, and M.S. Zhan, Entangling Two Individual Atoms of Different Isotopes via Rydberg Blockade, Phys. Rev. Lett. {\bf 119}, 160502 (2017).












\end{thebibliography}

\end{document}